\documentclass{article}
\usepackage[english]{babel}
\usepackage{amsfonts,amssymb, amsmath}

\newtheorem{prop}{Proposition}

\begin{document}


\title{On tensor invariants of the Clebsch system}
\author{A.V. Tsiganov\\
\it\small Steklov Mathematical Institute of Russian Academy of Sciences, Russia.\\
\it\small e--mail: andrey.tsiganov@gmail.com}
\date{}
\maketitle

\begin{abstract}
We present some new Poisson bivectors that are invariants by the Clebsch system flow. Symplectic integrators on their symplectic leaves exactly preserve the corresponding Casimir functions, which have different physical meanings. The Kahan discretization of the Clebsch system is discussed briefly.
\end{abstract}

\section{Introduction}
 \setcounter{equation}{0}
 
As engineering and science become increasingly complex, there is a growing need for more sophisticated numerical methods to support model-based design and analysis. The advent of enhanced computing capabilities enables the resolution of increasingly intricate problems over extended periods of time. The emergence of computational methods has led to the identification of fundamental properties such as accuracy, stability, convergence and computational efficiency as being of paramount importance in determining the efficacy of a numerical algorithm.

 In recent developments, various aspects of structure preservation have come to the fore as a significant addition to these fundamental properties \cite{book1,book2,erg01}. A fundamental principle underpinning the structure-preserving approach is the treatment of the numerical method as a discrete dynamical system, thereby approximating the flow of the governing continuous differential equation. This approach eschews the focus on numerical approximation of a single trajectory, instead emphasising the broader context of structure-preservation. This approach facilitates a more profound comprehension of the invariants and qualitative properties of the numerical method. It is notable that mechanical systems frequently manifest physically salient invariants, such as momentum, energy, vorticity, etc. The behaviour of these invariants during simulations serves as a crucial metric for evaluating the accuracy of the numerical method. However, most traditional numerical methods do not account for the underlying geometric structure of the physical system, thereby introducing numerical dissipation and failing to preserve the system's invariants. In contrast, a structure-preserving numerical method ensures that qualitative features, such as invariants of motion or the structure of the state space, are reflected in the simulation \cite{cos23,cos23a}. Furthermore, it can provide accurate numerical simulation over exponentially long times \cite{tao16,tao21}.

 Deep learning technology selects the most appropriate method of solving a given system of differential equations based on the existing experience of solving similar systems \cite{lut19,goy23,fang24,dias24}. The next generation of artificial intelligence technologies is aimed at automatic development of an efficient and stable numerical method of integration for each given system \cite{dier23,vac24}. This raises the question of constructing invariant geometric structures, or tensor invariants, for a given system of differential equations and discrete mappings that preserve these tensor invariants. At this stage, our research is focused on the automation of the computation of tensor invariants for a range of systems of equations arising in various applications.
 
Let us consider an autonomous system of ordinary differential equations 
\begin{equation}\label{m-eq}
\frac{d}{dt} {x} = X(x_1,\ldots,x_n)\,,
\end{equation} 
on the state space  with coordinates $x=(x_1,\ldots,x_n)$. Prior to the development of a structure-preserving integrator for (\ref{m-eq}), it is necessary to define the underlying  geometric structures $T$ that have to be preserved. These underlying invariant structures are solutions of the  invariance equation
\begin{equation} \label{g-inv}
\mathcal L_X\, T=0
\end{equation} 
on the tensors invariants  $T$ to the flow generated by the vector field $X=(X^1,\ldots,X^n)$,  including state space functions  (first integrals), multivector fields (symmetry fields, Poisson structures),  differential forms (symplectic form, volume form),  etc. See \cite{koz92,koz13,koz14,koz19,rat24} and references therein for a modern discussion of the relationship between equations (\ref{m-eq}) and (\ref{g-inv}).

The Lie derivative $L_X T$ determines the rate of change of the tensor field $T$ under the state space deformation defined by the flow of the system (\ref{m-eq}). In local coordinates the Lie derivative of the tensor field $T$ of type $(p, q)$ is equal to
\begin{align}
({\mathcal {L}}_{X}T)^{i_{1}\ldots i_{p}}_{j_{1}\ldots j_{q}}=\sum_{k=1}^n X^{k}(\partial _{k}T^{i_{1}\ldots i_{p}}_{j_{1}\ldots j_{q}})
&-\sum_{\ell=1}^n (\partial _{\ell}X^{i_{1}})T^{\ell i_{2}\ldots i_{p}}_{j_{1}\ldots j_{s}}-\ldots - \sum_{\ell=1}^n (\partial _{\ell}X^{i_{p}})T^{i_{1}\ldots i_{p-1}\ell}_{j_{1}\ldots j_{s}}\label{lie-d}\\
&+\sum_{m=1}^n(\partial _{j_{1}}X^{m})T^{i_{1}\ldots i_{p}}_{m j_{2}\ldots j_{q}}+\ldots +\sum_{m=1}^n(\partial _{j_{q}}X^{m})T^{i_{1}\ldots i_{p}}_{j_{1}\ldots j_{q-1}m}
\nonumber
\end{align}
where $\partial _k= {\partial }/{\partial x_k}$ is the partial derivative on the $x_k$ coordinate \cite{nt05}. 
 
Solutions of invariance equation (\ref{g-inv}) can be computed with the method of undetermined coefficients. In other words, we can use polynomial substitution for entries of tensor invariant $T$ that gives a polynomial system in the unknown coefficients of $T$. This approach can give rise to the computation of an exponential number of reducible tensor invariants and, therefore, we have to narrow the solution space and compute only some basic invariants \cite{ts24,ts25}.
 
In the  method of undetermined coefficients we do not use Lagrangian or Hamiltonian form of equations, symplectic or Poisson geometry, qualitative and asymptotic analysis of differential equations, Lax matrices, representation theory, etc.  We start with  a given vector field and  solve algebraic equations which are the backbone of many machine learning algorithms and techniques and, therefore, now we have many modern computer implementations of the desired  computational methods. 

As a benchmark model we chose the Clebsch system which the entries of $X$ are homogeneous polynomials of second order \cite{bm}. In 1892 K\"{o}otter solved Clebsch's equations of motion in terms of theta functions \cite{kot92}, which allows us to compare the numerical and analytical solutions. Later Bobenko obtained this analytical solution using finite-gap theory and $2\times 2$ Lax matrices \cite{bob87} whereas Zhivkov and Christov reproduce theta-functions formulae using  $4\times 4$ Lax matrices  \cite{ziv98}.

\section{Known tensor invariants for the Clebsch system}
 \setcounter{equation}{0}
The motion of a rigid body in an ideal fluid were obtained as Lagrangian equations of the second kind, or equations in generalized Lagrange coordinates by Kirchhoff in \cite{kirch}, see also \cite{tom}. The
Kirchhoff equations were rewritten in the Hamiltonian form by Clebsch \cite{cleb}. He also found a case of integrability of these equations in quadratures.

Among other forms of ordinary differential equations, the Lagrangian representation stands out due to its significant physical underpinnings and, building upon this framework, we can introduce the concept of Lagrangian neural networks, see \cite{lut19,dias24} and references therein. Hamiltonian neural networks are state-of-the-art models that regress the vector field of a dynamical system under the learning bias of Hamilton’s equations, see \cite{fang24,dier23,vac24,matt24}. 
Neural networks that synergistically integrate data and physical laws offer great promise in modeling dynamical systems. However, iterative gradient-based optimization of network parameters is often computationally expensive and suffers from slow convergence. We can accelerate the training of neural networks for approximating Hamiltonian systems through data-agnostic and data-driven algorithms. 

In \cite{put24,poi22} authors consider also the so-called Poisson neural networks. By definition Hamiltonian system contains two essential invariants: the Poisson bracket and the Hamiltonian. Hamiltonian systems with symmetries, whose paradigm examples are the Lie-Poisson systems, describe a broad category of physical phenomena, from satellite motion to underwater vehicles, fluids, geophysical applications, complex fluids, and plasma physics. The Poisson bracket in these systems comes from the symmetries, while the Hamiltonian comes from the underlying physics. In framework of the Lie-Poisson neural networks the Lie-Poisson bracket is known exactly, whereas the Hamiltonian is regarded as coming from physics and is considered not known, or known approximately.

Symbolic AI, which translates implicit human knowledge into a more formalised and declarative form based on rules and logic, should automatically develop its own structure-preserving method for each system of differential equations under consideration. Thus, the form of the equations is not essential and all the tensor invariants must be computed directly from (\ref{g-inv})
\[
\mathcal L_X T=0\,.
\] 
Input is a system of ordinary differential equations, nothing else. We need to obtain tensor invariants, find the basis of the invariants, and propose a discretization scheme that preserves these basic invariants exactly or approximately.

So, the input is the Kirchhoff equations for the Clebsch system which have the form  (\ref{m-eq}), 
where 
\[x=(p_1,p_2,p_3,M_1M_2M_3)\] 
and $X$ is a vector field   
\begin{equation}\label{x-cl}
X=\left(\begin{array}{c}
a_2M_2p_3 - a_3M_3p_2 \\
a_3M_3p_1-a_1M_1p_3 \\ 
a_1M_1p_2 - a_2M_2p_1 \\
(a_2 - a_3)(a_1p_2p_3 + M_2M_3) \\
(a_3 - a_1)(a_2p_1p_3 + M_1M_3) \\
(a_1 - a_2)(a_3p_1p_2 + M_1M_2) 
\end{array}\right)\,,
\qquad a_1,a_2,a_3\in\mathbb R\,,\\
\end{equation}
with entries which are homogeneous  polynomials of second order in $x$.

Here $M=(M_1,M_2,M_3)$ and $p=(p_1,p_2,p_3)$ are
three-dimensional vectors of "impulsive momentum" and "impulsive force" in the so-called body frame  \cite{kirch,cleb}.
Vector $\omega=(a_1M_1,a_2M_2,a_3M_3)$ is a vector of angular velocity and constants $ a_1,a_2$ and $a_3$ are defined by the attached masses and moments of inertia of the given body in the fluid. A full description of the long history of  this system can be found in the book of Borisov and Mamaev  \cite{bm}.

\subsection{Linear and quadratic tensor invariants}
In the space of scalar fields $f$ of type $(0,0)$ the invariance equation has the form
\[
\mathcal L_X f= X^1\frac{\partial f}{\partial x_1}+\cdots+X^6\frac{\partial f}{\partial x_6}=0\,.
\]
It is easy to solve this equation substituting  polynomials $f(x)$ with degree smaller some $N$ and solving the resulting system of linear equations on undetermined coefficients of this polynomial. The main problem here is to define the degree $N$ of the irreducible polynomials $f$, known as the Poincar\'{e} problem \cite{poi91}, for the given vector field $X$. 

For the Clebsch system as a base of solutions we take polynomials
\begin{align}
f_1=&p_1^2 + p_2^2 + p_3^2\,,\qquad f_2=p_1M_1+p_2M_2+p_3M_3\,,
\nonumber\\
f_3=&\frac{1}{2}\left(a_1M_1^2 + a_2M_2^2 + a_3M_3^2- a_2a_3p_1^2- a_1a_3p_2^2 -a_1a_2p_3^2\right)
=\frac{1}{2}(M,\omega)-V(p)\,,
\label{s-inv}\\
f_4=&\frac{1}{2}\left(M_1^2 + M_2^2 + M_3^2+a_1p_1^2 + a_2p_2^2 + a_3p_3^2 \right)=\frac{1}{2}M^2+U(p)\,.
\nonumber
\end{align}
Scalar invariants give rise to invariant one-forms $df_k$, i.e tensor invariants of type $(0,1)$, and tensor products of these one forms are also solutions of the equation (\ref{g-inv}).  

From the underlying physics we know that $f_1$ and $f_2$  are so-called the integrals of the impulsive force and impulsive momentum, respectively, whereas $f_3$  is a mechanical energy of the system \cite{kirch,cleb}. Unlike the Euler-Poisson equations, the value of the integral $f_1$, which expresses the invariance of the magnitude of the impulsive force, is not necessarily equal to unit.

In the space of the vector fields $Y$ of type $(1,0)$ the invariance equation (\ref{g-inv}) reads as
\[\mathcal L_X Y=[X,Y]=0\]
and using method of undetermined coefficients for entries of $Y$ we obtain invariant vector field
\begin{equation}\label{y-cl}
Y=\left(
    \begin{array}{c}
      p_3M_2 - p_2M_3 \\
      p_1M_3-p_3M_1   \\
      p_2M_1 - p_1M_2 \\
      (a_2 - a_3)p_2 p_3  \\
      (a_3 - a_1)p_1 p_3  \\
      (a_1 - a_2)p_1 p_2 
    \end{array}
  \right)
\end{equation} 
commuting with $X$ which entries are homogeneous  polynomials of second order in variables $x$.

In the space of multivector fields $P$ of valency $(2,0)$ the invariance equation (\ref{g-inv}) looks like
\begin{equation}\label{lie-2}
(L_X P)^{ij}=\sum_{k=1}^6\left(X^k\frac{\partial P^{ij}}{\partial x^k}-P^{kj}\frac{\partial X^i}{\partial x^k}-P^{ik}\frac{\partial X^j }{\partial x^k} \right)=0\,,\qquad 1\leq i,j\leq 6\,.
\end{equation}
Substituting multivector field with unknown coefficients
\[
P^{ij}=\sum_{k=1}^n c^{ij}_k x_k\,, \quad c^{ij}_k\in \mathbb R\,,\qquad 1\leq i,j \leq 6\,,
\]
into (\ref{lie-2}) and solving the resulting system of linear algebraic equations we obtain the following generic solution
\begin{equation}\label{g-lin}
P=b_1 P_1+b_2P_2\,,\qquad b_1,b_2\in \mathbb R\,,
\end{equation}
where
\begin{equation}\label{p1}
P_1=\left(
\begin{array}{cccccc}
0& 0& 0& 0& p_3& -p_2\\
0& 0& 0& -p_3& 0& p_1\\0& 0& 0& p_2& -p_1& 0
\\0& p_3& -p_2& 0& M_3& -M_2
\\-p_3& 0& p_1& -M_3& 0& M_1
\\p_2& -p_1& 0& M_2& -M_1& 0
\end{array}\right)
\end{equation}
and
\begin{equation}\label{p2}
P_2=\left(
\begin{array}{cccccc}0&M_3&-M_2&0&-a_1 p_3&a_1 p_2 \\
 -M_3&0&M_1&a_2p_3&0&-a_2p_1 \\ 
 M_2&-M_1&0&-a_3p_2 &a_3p_1 &0 \\
  0&-a_2p_3&a_3p_2&0&-a_3M_3 &a_2M_2\\ 
  a_1 p_3&0&-a_3p_1&a_3M_3&0&-a_1 M_1 \\ 
  -a_1 p_2&a_2p_1&0&-a_2M_2&a_1 M_1&0
\end{array}\right)\,.
\end{equation}
Here $P_1$ and $P_2$ are two compatible Poisson bivectors \cite{ts07,ts07a}, i.e. skew-symmetric tensor field $P$  (\ref{g-lin}) of valency $(2,0)$  satisfies to the Jacobi condition
\begin{equation}\label{jac-cond}
 [\![P,P]\!]=0\,,
 \end{equation}
 where $ [\![.,.]\!]$ is the Schouten-Nijenhuis bracket \cite{nt05}.  These bivectors determine two compatible Poisson brackets
 \[
 \{f,g\}_{1,2}=\sum_{i,j=1}^6 P^{ij}_{1,2}\frac{\partial f}{\partial x_i}\frac{\partial g}{\partial x_j}\,.
 \]
 In local coordinates, the rank of a Poisson structure $P$ is the rank of the matrix $P^{ij}$, such that $\mathrm{rank} P_{1,2}=4$ at the generic point $x$ by generic $a_{k}$.
 
 Let us underline that there are no symmetric solutions to the invariance equation (\ref{g-inv}) and its generic solution automatically satisfies the Jacobi condition.

\begin{prop}
Invariant Poisson bivector $P_1$ (\ref{p1}) is a Lie derivative of the invariant Poisson bivector $P_2$ (\ref{p2}) 
\[
P_1=\mathcal L_V P_2
\]
along the vector field $V$
\[
V=\frac{1}{2a_1a_2a_3}\,\left(
    \begin{array}{c}
    a_2 a_3 p_1\\ a_1 a_3 p_2\\ a_1 a_2 p_3\\ (a_2 + a_3) a_1M_1\\ (a_1 + a_3) a_2M_2 \\ (a_1 + a_2)a_3 M_3
     \end{array}
  \right)\,.
\]
It means that $P_1$ is a trivial deformations of $P_2$ in the Lichnerowicz-Poisson cohomology \cite{vai90}. The corresponding
Liouville vector field $V$ is not invariant to the flows associated with $X$ and $Y$. 
\end{prop}
The proof consists in solving linear algebraic equations on undetermined coefficients $c^j_k$ in polynomial substitution for entries of the Liouville vector field $V^j=\sum_{k} c^j_k x_k$ which to equations
\[
\mathcal L_{[X,V]}\,P_2=0\qquad\mbox{and}\qquad \mathcal L_{[Y,V]}\,P_2=0\,.
\]
Here $[.,.]$ is a Lie bracket, $X$ (\ref{x-cl}) and $Y$ (\ref{y-cl}) are invariant vector fields for the Clebsch system.

By definition tensor product of invariant bivectors with any invariant 1-form is invariant vector field
\[
{P}_{j}df_k=\alpha_{jk}X+\beta_{jk}Y\,, \qquad j=1,\ldots,6\,,\quad k=1,2,3,4\,,
\] 
which is a sum of two basic invariant vector fields $X$ and $Y$ with coefficients $\alpha_{jk}$ and $\beta_{jk}$ depending on the scalar invariants $f_1,\ldots,f_4$ (\ref{s-inv}). For instance, we have 
\[
P_1df_1=0\,,\qquad P_1df_2=0\,,\qquad P_1df_3=X\,,\qquad P_1df_4=Y
\]
and
\[
P_2df_1=-2Y\,,\qquad P_2df_2=0\,,\qquad P_2df_3=0\,,\qquad P_2df_4=X-(a_1+a_2+a_3)Y\,.
\]

Now let us consider  multivector field of valency $(2,0)$ with entries which are polynomials of second order with undetermined coefficients. Substituting multivector field $P$ with entries 
\[
P^{ij}=\sum_{k,m=1}^n c^{ij}_{km} x_kx_m\,, \quad c^{ij}_k\in \mathbb R\,,\qquad 1\leq i,j\leq 6\,,
\]
 into (\ref{lie-2}) and solving the resulting system of linear algebraic equations we obtain only trivial solution $P=0$.

Because divergency of $X$ is equal to zero
\[
\mbox{div}X=\frac{\partial X^1}{\partial x_1}+\cdots+\frac{\partial X^6}{\partial x_6}=0\,,
\]
then the Lie derivative of completely antisymmetric unit tensor fields $\Omega$ of type $(0,6)$ and $\mathcal E$ of type $(6,0)$ is zero
\begin{equation}\label{vol-inv}
 \Omega =dx_1\wedge dx_2\wedge\cdots\wedge x_6\,,
\qquad \mathcal L_X\Omega=0\,,
\end{equation}
and
\begin{equation}\label{lc-t}
 \mathcal E= \partial_1\wedge\partial_2\wedge\cdots\wedge\partial_6\,,\qquad
 \mathcal L_X \mathcal E=0\,,
\end{equation}
where $\partial _k= {\partial }/{\partial x_k}$ is the partial derivative on the $x_k$ coordinate. It enables us to compute another invariant skew symmetric tensor field with valence $(2,0)$
\begin{equation}\label{pf}
P_f=\mathcal E df_1df_2df_3df_4\,,\qquad \mbox{rank}\,P_f=2\,.
\end{equation}
Using the tensor product of the invariants $X$ and $Y$, we can also obtain the invariant tensor field of the valence $(2,0)$
\[
P_{XY}=X\cdot Y\,,
\] 
which can be divided on symmetric and skew symmetric parts. Entries of tensor invariants $P_f$ and $P_{XY}$ are polynomials of fourth order in coordinates $x$.

\section{Cubic invariant bivectors}
 \setcounter{equation}{0}
The main our aim is to describe  solutions of the invariance equation  in the space of multivector fields of valency $(2,0)$ with entries
\begin{equation}\label{cub-gen}
P^{ij}=\sum_{k,\ell,m=1}^6 c^{ij}_{k\ell m} x_kx_\ell x_m\,, \qquad c^{ij}_{k\ell,m}\in \mathbb R\,,\qquad 1\leq i,j\leq 6\,,
\end{equation}
which are homogeneous polynomials of third order in $x$. This multivector field is substituted into equation (2.3), the resulting system of linear algebraic equations is solved, and all the obtained solutions are described.

Since we are only interested in the invariance of this tensor field and the existence of a similar invariant for the discrete analog of a given system equations (\ref{m-eq}), we do not impose any additional conditions such as skew-symmetry or Jacobi identity.

\subsection{Invariant rational Poisson bivectors}
Let us begin by considering tensor fields (\ref{cub-gen}) which are invariant with respect to a pair of phase flows generated by vector fields $X$ (\ref{x-cl}) and $Y$ (\ref{y-cl}).
\begin{prop}
The general solution of a pair of invariance equations
\[
\mathcal{L}_XP=0\qquad\mbox{and}\qquad \mathcal{L}_YP=0
\]
in the space of tensor fields of valence $(2,0)$ with entries (\ref{cub-gen})
depends on 10 arbitrary parameters
\begin{equation}\label{p-xy}
P_{xy}=(c_1f_1+c_2f_2+c_3f_3+c_4f_4)P_1+
(c_5f_1+c_6f_2+c_7f_3+c_8f_4)P_2+
c_9\hat{P}_3+c_{10}\hat{P}_4\,.
\end{equation}
Here $P_1$ and $P_2$ are invariant Poisson bivectors (\ref{p1}-\ref{p2}) and $f_1,\ldots,f_4$ are scalar invariants (\ref{s-inv}).
\end{prop}
The proof consists in solving 3659 algebraic equations with respect to 840 coefficients $c^{ij}_{k\ell,m}$ included in the definition of cubic polynomials $P_{ij}$ (\ref{cub-gen}) and finding a linearly independent basis in the solution space.

The rank four invariant bivector $\hat{P}_3$ in (\ref{p-xy}) is equal to
 \begin{align}
\hat{P}_3=f_2&\left(\begin{smallmatrix}
0&   M_3&  - M_2&  0&(a_1 + a_3)p_3 & -(a_1+a_2)p_2  \\
*&  0&   M_1& -(a_2 + a_3)p_3 &  0& (a_1 + a_2)p_1 \\
*&  *&  0& (a_2 + a_3)p_2 & -(a_1 + a_3)p_1 &  0\\
*&  *& *&  0&  2a_3M_3& -2a_2M_2\\
*&  *& *&  *&  0&  2a_1M_1\\
*& *&  *&  *&  *&  0
\end{smallmatrix}\right)
\label{ph3}
\\
\nonumber\\
+&\left(
\begin{smallmatrix}
0& -p_3 M^2& p_2 M^2& (a_3-a_2) (Mp-2p_2 p_3M_1)& (a_3-a_1)  (p_2^2+p_3^2)M_3& (a_1-a_2)  (p_2^2+p_3^2)M_2\\
*& 0& -p_1 M^2& (a_2-a_3)  (p_1^2+p_3^2)M_3& (a_1-a_3) (Mp-2p_1 p_3M_2)& (a_1-a_2) (p_1^2+p_3^2)M_1 \\
*& *& 0& (a_2-a_3)  (p_1^2+p_2^2)M_2& (a_3-a_1)  (p_1^2+p_2^2)M_1& (a_2-a_1) (Mp-2p_1 p_2M_3)\\
*& *& *& 0&  (a_2-a_3) (a_3-a_1)(p_1^2+p_2^2) p_3&  (a_2-a_3) (a_2-a_1) (p_1^2+p_3^2)p_2\\
*& *& *& *& 0&  (a_3-a_1) (a_1-a_2)(p_2^2+p_3^2)p_1 \\
*&*&*&*&*& 0
\end{smallmatrix}\right)\,,\nonumber
\end{align}
where
\[
M^2=M_1^2+M_2^2+M_3^2\qquad\mbox{and}\qquad Mp=p_2 p_3M_1+p_1 p_3M_2+p_1 p_2M_3\,.
\]
\begin{prop}
 Using invariant  polynomial bivector (\ref{ph3}) we can define rational invariant Poisson bivector 
\begin{equation}\label{p3}{P}_3=f_2^{-1}\hat{P}_3,\qquad \mbox{rank}\,{P}_3=4,
\end{equation} 
which is compatible with the Poisson bivector $P_1$ (\ref{p1}), i.e. 
\[ 
[\![{P}_3, {P}_3]\!]=0\,,\qquad  [\![P_1, {P}_3]\!]=0\,.
\]
\end{prop}
Tensor products of this invariant Poisson bivector with invariant 1-forms $df_k$ are the following invariant combination of $X$ (\ref{x-cl}) and $Y$ (\ref{y-cl})
\[\begin{array}{lll}
&P_3df_1=- 2\,Y\,,\qquad
&P_3df_2=-\frac{f_1}{f_2}\,X +\frac{(a_1 + a_2 + a_3)f_1-2f_4}{f_2}\,Y\,,
\\ \\
&P_3df_3=2(a_1a_2 + a_1a_3 + a_2a_3)\,Y\,,\qquad 
&P_3df_4=-X +(a_1 + a_2 + a_3)\,Y\,.
\end{array}
\]
In the following we will use these relations to compute the Casimir functions of this rational Poisson bivector.

The rank four invariant bivector $\hat{P}_4$ in (\ref{p-xy})  has the following entries
\begin{align}
 \hat{P}_4^{12}=&\phantom{-}a_3M_3 (f_2 + p_3M_3) + p_3 (a_1M_1^2 + a_2M_2^2 )\,,\nonumber\\
 \hat{P}_4^{13}=&-a_2M_2 (f_2 + p_2M_2) - p_2 (a_1M_1^2  + a_3M_3^2 )\,,\nonumber\\
  \hat{P}_4^{23}=&\phantom{-}a_1M_1 (f_2+ p_1M_1)  + p_1 (a_2M_2^2  + a_3M_3^2 )\,,
 \nonumber\\
 \hat{P}_4^{14}=&(a_2 - a_3)(-a_1M_1p_2 p_3 + a_2M_2 p_1 p_3 +a_3 M_3 p_1 p_2 + M_1 M_2 M_3)\,,
\nonumber\\
\hat{P}_4^{25}=&(a_3 - a_1)(\phantom{-} a_1M_1 p_2 p_3 - a_2M_2 p_1 p_3 + a_3M_3 p_1 p_2 + M_1 M_2 M_3)\,,
\nonumber\\
\hat{P}_4^{36}=&(a_1 - a_2)(\phantom{-}a_1M_1 p_2 p_3 + a_2M_2 p_1 p_3 - a_3M_3 p_1 p_2 + M_1 M_2 M_3)\,.
\label{ph4}\\
  \hat{P}_4^{15}=&-2 M_3 a_1 a_3 p_3^2 - p_3 (f_2-p_3M_3) (a_1 a_2 + 2 a_1 a_3 - a_2 a_3) - M_3 (M_1^2-a_3 p_2^2 ) (a_1 - a_3)\,,\nonumber\\
 \hat{P}_4^{16}=&\phantom{-}2 M_2 a_1 a_2 p_2^2 + p_2 (f_2-p_2M_2) (2 a_1 a_2 + a_1 a_3 - a_2 a_3) + M_2 (M_1^2-a_2 p_3^2 ) (a_1 - a_2)\,,\nonumber\\
 \hat{P}_4^{24}=&\phantom{-}2 M_3 a_3 a_2 p_3^2 + p_3(f_2-p_3M_3) (a_1 a_2 - a_1 a_3 + 2 a_2 a_3)  + M_3 (M_2^2-a_3 p_1^2 +M_2^2) (a_2-a_3)\,,\nonumber\\
 \hat{P}_4^{26}=&-2 a_1 M_1 a_2 p_1^2 - p_1(f_2-p_1M_1) (2 a_1 a_2 - a_1 a_3 + a_2 a_3) + M_1 ( M_2^2-a_1 p_3^2 ) (a_1 - a_2)\,,\nonumber\\
 \hat{P}_4^{34}=& -2 a_3 M_2 a_2 p_2^2 + p_2(f_2-p_2M_2) (a_1 a_2 - a_1 a_3 - 2 a_2 a_3) + M_2 (M_3^2-a_2 p_1^2 ) (a_2-a_3)\,,\nonumber\\
 \hat{P}_4^{35}=&\phantom{-}2 a_3 M_1 a_1 p_1^2 - p_1(f_2-p_1M_1) (a_1 a_2 - 2 a_1 a_3 - a_2 a_3)  - M_1 (M_3^2-a_1 p_2^2 ) (a_1 - a_3)\,,\nonumber\\
 \hat{P}_4^{45}=&-2p_3 a_3^2 M_3^2 + (a_1 a_2 - a_1 a_3 - a_2 a_3 - a_3^2) (f_2-p_3M_3) M_3 + p_3 (a_1 p_2^2 + a_2 p_1^2) (a_2-a_3 ) (a_1 - a_3)\,,\nonumber\\
\hat{P}_4^{46}=&\phantom{-}2 p_2 a_2^2 M_2^2 + (a_1 a_2 - a_1 a_3 + a_2^2 + a_2 a_3) (f_2-p_2M_2) M_2 + p_2 (a_1 p_3^2 + a_3 p_1^2) (a_2-a_3) (a_1 - a_2)\,,\nonumber\\
\hat{P}_4^{56}=&-2 p_1 a_1^2 M_1^2 - (a_1^2 + a_1 a_2 + a_1 a_3 - a_2 a_3) (f_2-p_1M_1) M_1 + p_1 (a_2 p_3^2 + a_3 p_2^2) (a_1 - a_3) (a_1 - a_2)\,.\nonumber
\end{align}
As above this invariant polynomial bivector can be reduced to the rational bivector which satisfies to the Jacobi condition.
\begin{prop}
Using invariant polynomial bivector (\ref{ph4}) we define rational Poisson bivector 
\begin{equation}\label{p4}
{P}_4=f_2^{-1}\hat{P}_4,,\qquad  \mbox{rank}\,P_4=4\,,
\end{equation}
which is compatible with the Poisson bivector $P_2$ (\ref{p2}), i.e. 
\[ 
[\![\hat{P}_4, \hat{P}_4]\!]=0\,,\qquad [\![P_2, \hat{P}_4]\!]=0\,.\]

\end{prop}
Tensor products of this invariant Poisson bivector with invariant 1-forms $df_k$ are equal to
\begin{align*}
&P_4df_1=-2X\,,\\ &P_4df_2=\frac{(a_1 + a_2 + a_3)f_1-2f_3}{f_2}\,X + \frac{4f_3}{f_2}\, Y\,,\\
&P_4df_3=(a_1a_2 + a_1a_3 + a_2a_3)\,X - 4a_1a_2a_3\,Y\,,\\  
&P_4df_4=(a_1 + a_2 + a_3)\,X + 2(a_1a_2 + a_1a_3 + a_2a_3)\,Y\,.
\end{align*}

Summing up, invariant polynomial bivectors $\hat{P}_3$ and $\hat{P}_4$  define  a pair of rational Poisson brackets  compatible with linear invariant Poisson brackets associated with  bivectors $P_1$ and $P_2$. 

\subsection{Invariant polynomial Poisson bivectors}
Now let us consider tensor field (\ref{cub-gen}) which is invariant with respect to one of the flows generated by  vector fields $X$ (\ref{x-cl}) or $Y$ (\ref{y-cl}).
\begin{prop}
The general solution of the invariance equations
\begin{equation}\label{p-x}
\mathcal{L}_XP=0
\end{equation}
in the space of tensor fields of valence $(2,0)$ with entries (\ref{cub-gen})
depends on 11 arbitrary parameters
\begin{equation}\label{p-cub-x}
P_x=P_{xy}+c_{11}P_5\,,
\end{equation}
where $P_{xy}$ is given by (\ref{p-xy}).
\end{prop}
The proof consists in solving linear algebraic equations on undetermined coefficients.

The rank four invariant bivector $P_5$ in (\ref{p-x}) is given by 
\begin{equation}\label{p5}
P_5=\hat{P}_5 + (2f_4 - (a_1 + a_2 + a_3)f_1)P_2\,,\qquad \mbox{rank}\,P_5=4,
\end{equation}
where
\begin{align*}
\hat{P}_5^{12}&=\phantom{-}2a_3(f_1-p_3^2)M_3 -2 (a_1p_1M_1+a_2p_2M_2)p_3\,,\nonumber\\
\hat{P}_5^{13}&=-2a_2(f_1-p_2^2)M_2 + 2(a_1p_1M_1 + a_3p_3M_3)p_2\,,\nonumber\\
\hat{P}_5^{23}&=\phantom{-}2a_1(f_1-p_1^2)M_1 -2 (a_2p_2M_2+a_3p_3M_3)p_1\,,\nonumber\\
\hat{P}_5^{14}&=\phantom{-}2(a_2 - a_3) (a_1 p_2 p_3 + M_2 M_3) p_1 - 2 M_1 (p_3a_2M_2 - p_2a_3M_3)\,,\nonumber\\
\hat{P}_5^{2,5}&=-2 (a_1 - a_3) (a_2 p_1 p_3 + M_1 M_3) p_2 + 2 M_2 (p_3 a_1 M_1 -p_1a_3 M_3)\,,\nonumber\\
\hat{P}_5^{3,6}&=\phantom{-}2 (a_1 - a_2) (a_3 p_1 p_2 + M_1 M_2) p_3 - 2 M_3(p_2a_1M_1 -p_1a_2 M_2)\,,\\
\hat{P}_5^{15}&=-2 a_2 \left((a_1- a_3) p_1^2 + M_2^2\right) p_3 - 2 M_3\left((a_1-a_3)p_1M_1- a_3 p_2M_2\right)\,,\nonumber\\
\hat{P}_5^{16}&=\phantom{-}2a_3\left((a_1- a_2) p_1^2 + M_3^2\right)p_2 + 2 M_2 \left((a_1-a_2)p_1 M_1  - a_2p_3M_3\right)\,,\nonumber\\
\hat{P}_5^{24}&=\phantom{-}2a_1\left((a_2- a_3)p_2^2 + M_1^2\right)p_3 + 2 M_3 \left((a_2-a_3)p_2M_2-a_3p_1 M_1\right)\,,\nonumber\\
\hat{P}_5^{26}&=-2a_3\left((a_2-a_1) p_2^2 + M_3^2\right) p_1 - 2 M_1 \left((a_2 -a_1) p_2 M_2  -a_1 p_3M_3\right)\,,\nonumber\\
\hat{P}_5^{34}&=-2a_1\left((a_3-a_2) p_3^2  + M_1^2\right) p_2 - 2 M_2 \left((a_3-a_2)p_3M_3- a_2p_1M_1 p_1\right)\,,\nonumber\\
\hat{P}_5^{35}&=\phantom{-}2 a_2\left( (a_3-a_1) p_3^2 + M_2^2\right) p_1 + 2 M_1 \left((a_3-a_1)p_3M_3-a_1p_2M_2\right)
 \end{align*}
 and
 \begin{align*}
 \hat{P}_5^{45}&=2\left((a_3-a_1) M_1^2 +(a_3-a_2) M_2^2\right) M_3 - 2\left(a_2(a_1 - a_3) p_1 M_1  + a_1(a_2-a_3)p_2 M_2\right)p_3\,,\\
 \hat{P}_5^{46}&=2\left((a_1-a_2) M_1^2 +(a_3-a_2)M_3^2\right) M_2 + 2 \left(a_3(a_1 - a_2) p_1 M_1  +a_1(a_3-a_2)p_3 M_3\right)p_2\,,\\
 \hat{P}_5^{56}&=2\left((a_1 -a_2)M_2^2 + (a_1-a_3) M_3^2\right) M_1 + 2\left(a_3(a_1-a_2)p_2M_2 + a_2(a_1-a_3) p_3M_3\right)p_1\,.
 \end{align*}
Derivatives of this bivector $P_5$ (\ref{p5}) are
\[\mathcal{L}_XP_5=0 \qquad\mbox{and}\qquad 
\mathcal L_Y P_5=P_f\,,
\]
where $P_f$ is a rank two invariant bivector (\ref{pf}) obtained from completely antisymmetric unit tensor fields $\mathcal E$ and scalar invariants $f_1,\ldots,f_4$ (\ref{s-inv}).

 Bivector $P_5$ satisfies to the Jacobi condition
\[
[\![P_5,P_5]\!]=0
\]
and compatibility conditions
\[
[\![f_2P_2,P_5]\!]=0\qquad\mbox{and}\qquad [\![f_3P_2,P_5]\!]=0\,.
\]
So, we have invariant polynomial Poisson bivector $P_5$ which is compatible with invariant bivectors $f_2P_2$ and $f_3P_2$.
Remind that $f_2$ and $f_3$ are the Cazimir elements of the Poisson bivector $P_2$
\[
P_2df_2=0\qquad \mbox{and}\qquad P_2df_3=0\,,
\]
and, therefore, $f_2P_2$ and $f_3P_2$ are Poisson bivectors which satisfy to the Jacobi condition.
 
Tensor products of invariant Poisson bivector $P_5$ with invariant 1-forms $df_k$ are the following invariant combination of $X$ (\ref{x-cl}) and $Y$ (\ref{y-cl}) 
 \begin{align*}
 &P_5df_1=- 4f_1X + 2\left( \left(a_1 + a_2 + a_3\right)f_1-2f_4\right)\,Y\,,\qquad
 P_5df_2=-4f_2\,X\,,\qquad
 P_5df_3=-4f_3X\,,\\ \\
 &P_5df_4=-\left( \left(a_1 + a_2 + a_3\right)f_1+2f_4\right)\,X + (a_1 + a_2 + a_3) \left( \left(a_1 + a_2 + a_3\right)f_1-2f_4\right)\,Y\,.
 \end{align*}
In the following we will use these relations to compute the Casimir functions of this cubic Poisson bivector.

\begin{prop}
The general solution of the invariance equations
\begin{equation}\label{p-y}
\mathcal{L}_YP=0
\end{equation}
in the space of tensor fields of valence $(2,0)$ with entries (\ref{cub-gen})
depends on 11 arbitrary parameters
\begin{equation}\label{p-cub-y}
P_y=P_{xy}+c_{11}P_6\,,
\end{equation}
where $P_{xy}$ is given by (\ref{p-xy}).
\end{prop}
The proof consists in solving linear algebraic equations on undetermined coefficients.

The rank four invariant bivector $P_6$ in (\ref{p-y}) has the following entries  
\begin{align}
P_6^{12}&=\phantom{-}2(p_3^2-f_1)M_3 + 2(f_2-p_3M_3)p_3\,,\nonumber\\ 
P_6^{13}&=-2(f_1-p_2^2)M_2 - 2(f_2-p_2M_2)p_2\,,\nonumber\\
P_6^{23}&=\phantom{-}2(p_1^2-f_1)M_1 + 2(f_2-p_1M_1)p_1\,,\nonumber\\
P_6^{14}&=\phantom{-}2(p_3M_2 - p_2M_3)M_1 - 2(a_2 - a_3)p_1p_2p_3\,,\nonumber\\
P_6^{25}&=-2(p_3M_1-p_1M_3)M_2 + 2(a_1 - a_3)p_1p_2p_3\,,\label{p6}\\
P_6^{36}&=\phantom{-}2(p_2M_1 - p_1M_2)M_3 - 2(a_1 - a_2)p_1p_2p_3\,,\nonumber\\
P_6^{45}&=\phantom{-}2\left((a_2 - a_3)p_2M_2 + (a_1 - a_3)p_1M_1\right)p_3 - 2f_4M_3\,,\nonumber\\
P_6^{46}&=\phantom{-}2\left((a_2 - a_3)p_3M_3 + (a_2 - a_1)p_1M_1\right)p_2+ 2f_4M_2\,,\nonumber\\
P_6^{56}&=\phantom{-}2\left((a_3 - a_1)p_3M_3 + (a_2 - a_1)p_2M_2\right)p_1 - 2f_4M_1\nonumber
\end{align}
and
\begin{align*}
P_6^{15}&=-a_3 p_3^3 + (\phantom{-}a_1 p_1^2 - a_2 p_2^2 - 2 a_3 p_1^2 - M_1^2 + M_2^2 - M_3^2) p_3 - 2 M_3 M_2 p_2\,,\\
P_6^{16}&=\phantom{-}a_2p_2^3 + (-a_1 p_1^2 + 2 a_2 p_1^2 + a_3 p_3^2 + M_1^2 + M_2^2 - M_3^2) p_2 + 2 M_3 M_2 p_3\,,\\ 
P_6^{24}&=\phantom{-}a_3 p_3^3 + (\phantom{-}a_1 p_1^2 - a_2 p_2^2 + 2 a_3 p_2^2 - M_1^2 + M_2^2 + M_3^2) p_3 + 2 M_3 M_1 p_1\,,\\
P_6^{26}&=-a_1p_1^3 + (-2 a_1 p_2^2 + a_2 p_2^2 - a_3 p_3^2 - M_1^2 - M_2^2 + M_3^2) p_1 - 2 M_3 M_1 p_3\,,\\
P_6^{34}&=-a_2p_2^3 + (-a_1 p_1^2 - 2 a_2 p_3^2 + a_3 p_3^2 + M_1^2 - M_2^2 - M_3^2) p_2 - 2 M_2 M_1 p_1\,,\\
P_6^{35}&=\phantom{-}a_1p_1^3 + (\phantom{-}2 a_1 p_3^2 + a_2 p_2^2 - a_3 p_3^2 + M_1^2 - M_2^2 + M_3^2) p_1 + 2 M_2 M_1 p_2\,.
\end{align*}
It has the following properties
\[
\mathcal L_X P_6=P_f\qquad\mbox{and}\qquad \mathcal{L}_YP_6=0\,,
\]
where $P_f$ is a rank two invariant bivector (\ref{pf}) obtained from completely antisymmetric unit tensor fields $\mathcal E$ and scalar invariants $f_1,\ldots,f_4$ (\ref{s-inv}). Moreover, $P_6$ satisfies to the Jacobi condition
\[
[\![P_6,P_6]\!]=0
\]
and compatibility conditions
\[
[\![f_1P_1,P_6]\!]=0 \qquad\mbox{and}\qquad [\![f_2P_1,P_6]\!]=0\,.
\]
So, we have invariant polynomial Poisson bivector $P_6$ which is compatible with invariant bivectors $f_1P_1$ and $f_2P_1$.
Remind that $f_1$ and $f_2$ are the Cazimir elements of the Poisson bivector $P_1$
\[
P_1df_1=0\qquad \mbox{and}\qquad P_1df_2=0\,,
\]
and, therefore, $f_1P_1$ and $f_2P_1$ are  the Poisson bivectors which satisfy to the Jacobi condition. 

Tensor products of this invariant Poisson bivector with invariant 1-forms $df_k$ are equal to
\[
P_6df_1=4f_1\,Y\,,\quad
P_6df_2=4f_2\,Y\,,\quad
P_6df_3=-2f_4\,X + 4f_3\,Y\,,\quad
P_6df_4=2f_4\,Y\,.
\]
 
Summing up, invariant Poisson bivectors $\hat{P}_5$ and $\hat{P}_6$  define  a pair of cubic Poisson brackets compatible with the Poisson brackets associated with invariant bivectors $f_{1,2}P_1$ and $f_{2,3}P_2$. 
 
 \subsection{Moser-Veselov discretization scheme}
Equations of motion for the Clebsch system can be expressed in a Lax form
 \begin{equation}\label{lax-g}
\frac{d}{dt}  L(\lambda)=[  L(\lambda),   A(\lambda)]\,,
\end{equation}
which automatically exhibits scalar invariants as eigenvalues of
$ L$ and leads to the linearization of the flow on the
Jacobi or Prym varieties of the algebraic curve $\det\left(
L(\lambda)-\mu\mathbf I\right)=0$ \cite{bob87, ziv98}. Here $\lambda$ is an auxiliary variable
(spectral parameter) and $[.,.]$ is a commutator of two matrices.
 
One of the Lax representations for the Clebsch system is related with
K\"otter's approach \cite{kot92}
\[L(\lambda)=\sum_{k=1}^3\left(w_k M_k+\frac{w_1w_2w_3}{w_k}p_k \right)\sigma_k
\,,\qquad{A}(\lambda)=\sum_{k=1}^3p_kw_k\sigma_k\,,
\]
where $\sigma_k$ are the Pauli matrices and $w_k=\sqrt{\lambda-a_k}$  can be considered as basic elliptic
functions, see details in \cite{bbe94,bob87}.
 
Another Lax representation for the Clebsch system was found
by Perelomov   \cite{p81}
\[
 L(\lambda)=
 \left(
   \begin{array}{ccc}
     \lambda^2p_1^2+a_1 & \lambda^2p_1p_2 + M_3\lambda  & \lambda^2p_1p_3 - M_2\lambda  \\
     \lambda^2p_1p_2 - M_3\lambda  & \lambda^2p_2^2+a_2 & \lambda^2p_2p_3 + M_1\lambda  \\
     \lambda^2p_1p_3 + M_2\lambda & \lambda^2 p_2p_3 - M_1\lambda  & \lambda^2p_3^2 +a_3
   \end{array}
 \right)\,,
\]
and
\[
A(\lambda)=2\lambda \left(
   \begin{array}{ccc}
    p_1^2 & p_1p_2   & p_1p_3  \\
     p_1p_2  & p_2^2& p_2p_3\\
     p_1p_3& p_2p_3   & p_3^2 
   \end{array}
 \right)\,.
\]
One more  Lax representation for the Clebsch system may be found in \cite{ziv98}. Two-dimensional family of $4\times 4$ Lax matrices could be also obtained from a family of Lax matrices for the Frahm-Schottky-Manakov system on $so(4)$ obtained by Adler and van Moerbeke \cite{avm88}. For brevity we do not reproduce this family of Lax matrices, which can be found in \cite{kt05}.

Using these Lax matrices we can construct discretization of the Clebsch system  by refactorization of matrix polynomials. This discretization is represented by an isospectral transformation
 \[
 L(\lambda)\to \widetilde{L}(\lambda)=A(\lambda) L(\lambda)  A^{-1}(\lambda)
 \]
 which does not explicitly involve a time step, see details in the Moser and Veselov paper  \cite{mos91}. We do not know whether this discrete mapping preserves the tensor invariants of the continuous system.
 
Six invariant bivectors $P_1,\ldots,P_6$  define six invariant Poisson brackets 
\[
\{f,g\}^{(k)}=\sum_{i,j=1}^6 P_k^{ij}\frac{\partial f}{\partial x_i}\frac{\partial g}{\partial x_j}\,,
\]
so that scalar invariants 
\[
I_n = {\rm Tr} \;(L^n(\lambda))
\]
are in the involution under all these  Poisson brackets. It is equivalent to the existence of
the six classical $r$-matrices associated with these brackets
\[
\{ L_1(\lambda) , L_2(\mu) \}^{(k)} = [r^{(k)}_{12}(\lambda,\mu), L_1(\lambda)] -[r^{(k)}_{21}(\lambda,\mu),L_2(\mu)]\,,
\]
where we use standard notation from \cite{bab90}. 

The Jacobi identity for bivectors $P_k$ 
\[
[\![P_k,P_k]\!]=0\,,\qquad k=1,\ldots,6\,,
\]
leads to equation
\[\{L_1, [r_{12}, r_{13}] + [r_{12}, r_{23}] + \{L_2, r_{13}\}- \{L_3, r_{12}\}\} + \mbox{cycl.perm.} = 0.\]
If the $r$-matrix is constant, then only the first term survives and the Jacobi identity is satisfied
if
\begin{equation}\label{c-rmat}
[r_{12}, r_{13}] + [r_{12}, r_{23}] + [r_{23}, r_{13}] = 0.
\end{equation}
If the $r$-matrix is antisymmetric $r_{12} = -r_{21}$ then the corresponding equation is called the classical
Yang-Baxter equation \cite{rs94}.

Linear Poisson bivector $P_1$ (\ref{p1}) yields a constant $r$-matrix that satisfies the classical Yang-Baxter equation. Second linear Poisson bivector $P_2$ (\ref{p2}) gives a constant $r$-matrix that satisfies (\ref{c-rmat}). As outlined in Proposition 1, linear invariant bivectors $P_1$ and $P_2$ are related to each other, and therefore the corresponding constant $r$-matrices are also related to each other. 

Associated with the other four bivectors $P_3, P_4, P_5$ and $P_6$, classical $r$-matrices depend on dynamical variables, similar to those for other families of classical integrable systems, see \cite{eekt94} and references therein.

In accordance with the definition, invariant Poisson bivectors satisfy to the invariance equation (\ref{g-inv})
\[
\mathcal L_X P_k=0\qquad\mbox{and}\qquad \mathcal L_YP_k=0\,,
\]
 which also leads to an existing  of the corresponding  $r$-matrices. It will be interesting to find a counterpart of these equations in the Moser and Veselov discretization scheme. 
 
\section{Poisson neural networks}
 \setcounter{equation}{0}
The symplectic integrators have been extended for systems with invariant Poisson bivectors  by assuming that the dynamics occurs within a neighbourhood where invariant Poisson structure has a constant rank that allows us to use local canonical coordinates according to the Lie-Darboux theorem \cite{lie,wei83}.. This method was called Poisson neural networks \cite{cos23, cos23a, put24, poi22}. This approach is used to develop a network based on transformations that exactly preserve the Poisson bracket and Casimir functions to machine precision.

\subsection{Symplectic leaves}
 For the Clebsch system we computed six invariant bivectors $P_1,\ldots,P_6$ using the generic polynomial substitution for entries of the multivector field  of valency $(2,0)$  in the equation (\ref{lie-2}). Every Poisson manifold is essentially
a union of symplectic manifolds which fit together in a smooth way. A Casimir function $C$ is constant along each symplectic leaf of the Poisson bivector $P$
 \[PdC=0\,,\]
and in a region where the rank of $P$ is constant, the symplectic leaves are exactly the common level manifolds of the Casimir functions.
 
For linear bivectors  $P_1$ (\ref{p1}) and $P_2$ (\ref{p2}) we have Casimir functions
 \[C_1^{(1)}=f_1\,,\qquad C_2^{(1)}=f_2\qquad\mbox{and}\qquad C_1^{(2)}=f_2\,,\qquad C_2^{(2)}=f_3 \]
For rational bivectors $P_3$ (\ref{p3}) and $P_4$ (\ref{p4}) the corresponding Casimir functions are linear and quadratic polynomials in the scalar invariants $f_k$ (\ref{f-int})
 \[
 C_1^{(3)}=(a_1a_2 + a_1a_3 + a_2a_3)f_1 + f_3\,,\qquad C_2^{(3)}=2f_1f_4 - f_2^2 \]
 and
 \begin{align*}
 &C_1^{(4)}(a_1^2a_2^2 + a_1^2a_3^2 + a_2^2a_3^2)f_1 + 2(a_1a_2 + a_1a_3 + a_2a_3)f_3 - 4a_1a_2a_3f_4\,,\\ 
 &C_2^{(4)}= (a_1a_2 + a_1a_3 + a_2a_3)^2f_1^2+ 4a_1a_2a_3f_2^2 + 4f_3^2 + 4\left(\left(a_1a_2 + a_1a_3 + a_2a_3\right)f_3 - 2a_1a_2a_3f_4\right)f_1\,.
   \end{align*}
The Casimir functions of cubic Poisson bivector $P_5$ (\ref{p5}) are rational functions
\[ 
 C_1^{(5)}=\frac{f_3}{f_2}\qquad\mbox{and}\qquad C_2^{(5)}=\frac{1}{f_2}\left(f_4 - \frac{(a_1 + a_2 + a_3)f_1}{2}\right)^2 \]
 similar to the Casimir functions for invariant Poisson bivector $P_6$ (\ref{p6})
 
 \[
 C_1^{(6)}=\frac{f_2}{f_1}\qquad\mbox{and}\qquad  C_2^{(6)}=\frac{f_4^2}{f_1}\,.
\]
These Casimir functions can be used to construct the Poisson  networks based on Lie-Darboux  transformations that exactly preserve these bivectors and their Casimir elements to machine precision. Which of these six networks is best for numerically studying the Clebsch system? In order to answer this question, a detailed comparison of the numerical performance of all these networks is required.

Invariant Poisson bivectors $P_1,\ldots,P_6$ are equivalent from a mathematical point of view. But the corresponding  symplectic leaves
\[
S^{(k)}_{e_1,e_2}=\{x\in\mathbb R^6|\quad C_1^{(k)}(x)=e_1\,,\quad C_2^{(k)}(x)=e_2\,\quad e_{1,2}\in \mathbb R\}
\]
 are different from a computational point of view. 

Symplectic integrator on the symplectic manifold $S^{(1)}_{e_1,e_2}$ preserves exactly 
integrals of the impulsive force $f_1=e_1$ and the impulsive momentum $f_2=e_2$, while the mechanical energy of the system $f_3$ and the invariant $f_4$ are approximately preserved. 

Symplectic integrator on symplectic manifold $S^{(2)}_{e_1,e_2}$  preserves exactly
integral of impulsive momentum $f_2=e_1$  and mechanical energy of the system $f_3=e_2$, while integrals of  impulsive force $f_1$ and invariant $f_4$ are approximately preserved. 

The combination of these two integrators could exactly preserve the linear Poisson bivectors $P_{1,2}$ and the three integrals of motion $f_1,f_2$ and $f_3$. If we want to add $f_4$ to this list, we have to consider symplectic leaves of rational or cubic bivectors $P_3,\ldots,P_6$. As an example, the symplectic integrator on the symplectic manifold $S^{(6)}_{e_1,e_2}$ preserves exactly the ratio of the integrals of the impulsive force and the impulsive momentum $f_2/f_1=e_1$ together with $f_4^2/f_1=e_2$, while the mechanical energy of the system is approximately preserved.

In a similar manner to the Runge-Kutta-Fehlberg method for the numerical solution of ordinary differential equations, we could construct a combination of symplectic integrators on various symplectic leaves to produce a few  estimates of different accuracy for scalar invariants $f_1,\ldots,f_4$, allowing for automatic error correction. 

\subsection{Darboux coordinates}
According to  Weinstein splitting theorem \cite{wei83} for a Poisson manifold 
there are a neighborhood $U$ of any point $x_0$ and  isomorphism $\varphi$ = $\varphi_S\times\varphi_N$
from $U$ to a product $S \times N$ such that $S$ is symplectic and the rank of $N$ at $\varphi_N(x_0)$ is zero. The factors $S$ and $N$ are unique up to local isomorphism.

An easy corollary of the existence part of the splitting theorem \cite{wei83} is the
following basic result of Lie \cite{lie}.  Suppose that the rank of the Poisson manifold is constant near
$x_0$. Then there are coordinates $(q_1, \ldots q_n,p_{q_1},\ldots,p_{q_k}, y_1,\ldots ,y_{s})$ near $x_0$ satisfying the canonical bracket relations
\[\{q_i, q_j\}= \{p_{q_i},p_{q_j}\}=\{q_i,y_j\}=\{p_{q_i},y_j\}=\{y_i,y_j\}=0\quad\mbox{and}\quad \{q_i,p_{q_j}\}= \delta_{ij}\,.\] 
The coordinates $(q, p_q)$ are known as Darboux coordinates on the symplectic leaf $S$, whilst $\varphi$ is referred to as the Lie-Darboux transformation.

On symplectic leaves of a Poisson bivector there are many different Darboux coordinates, which are related to each other by canonical transformations. For instance, on the symplectic leaves  of linear  bivector $P_1$ defined by
\[
f_1=e_1\,,\qquad f_2=e_2\,,
\]
 there are Euler variables, Andoyer variables, a counterpart of elliptic coordinates on the sphere, etc \cite{bm,ts22}. Let us consider the following  Lie-Darboux transformation $\varphi$  defined by equations \cite{sok10} 
\begin{align*}
M_1=&-{p_{q_1}} {q_1} {q_2}+\frac{{p_{q_2}} \left({q_1}^{2}-{q_2}^{2}-1\right)}{2}-\frac{{e_2} {q_1} \left({q_1}^{2}+{q_2}^{2}+1\right)}{2 {e_1} \left({q_1}^{2}+{q_2}^{2}\right)}\\
{M_2}=&\phantom{-} {p_{q_2}} {q_1} {q_2}+\frac{{p_{q_1}} \left({q_1}^{2}-{q_2}^{2}+1\right)}{2}-\frac{{e_2} {q_2} \left({q_1}^{2}+{q_2}^{2}+1\right)}{2 {e_1} \left({q_1}^{2}+{q_2}^{2}\right)}\\
 {M_3}=&\phantom{-} {p_{q_2}} {q_1}-{p_{q_1}} {q_2}
 \end{align*}
 and
 \[
 {p_1}= \frac{2 {e_1} {q_1}}{{q_1}^{2}+{q_2}^{2}+1}\,,\qquad
 {p_2}=\frac{2 {e_1} {q_2}}{{q_1}^{2}+{q_2}^{2}+1}\,,\qquad
 {p_3}= -\frac{{e_1} \left({q_1}^{2}+{q_2}^{2}-1\right)}{{q_1}^{2}+{q_2}^{2}+1}\,.
 \]
 In the canonical variables $q,p_q$ remaining pair of first integrals $f_3$ and $f_4$ (\ref{f-int}) have the so-called quasi-St\"{a}ckel form \cite{sok10}.   Let us present explicitly several equations of motion 
 \begin{equation}\label{x-s1}
 \begin{array}{lll}
 \{q_1,f_3\}_1=&\left(\frac{\left({q_1}^{2}-{q_2}^{2}+1\right)^{2} {a_2}}{4}+{q_2}^{2} \left({a_1} \,{q_1}^{2}+{a_3}\right)\right) {p_{q_1}}+\left(\frac{(q_1^2 - q_2^2 + 1)(a_2-a_1)}{2} - (a_1 - a_3)\right)q_1q_2p_{q_2}\\
 &+
 \frac{{e_2} {q_2} \left({q_1}^{2}+{q_2}^{2}+1\right) \left(2 {a_1} \,{q_1}^{2}-{a_2} \left(q_1^2 - q_2^2 + 1\right)\right)}{4 {e_1} \left({q_1}^{2}+{q_2}^{2}\right)}\,,
  \\ 
  \\
 \{q_2,f_3\}_1=&  \left(\frac{\left({q_1}^{2}-{q_2}^{2}-1\right)^{2} {a_1}}{4}+{q_1}^{2} \left({a_2} \,{q_2}^{2}+{a_3}\right)\right)p_{q_2}-
 \left(\frac{(q_1^2 - q_2^2 - 1)(a_1 - a_2)}{2} - (a_2 - a_3)\right)q_1q_2p_{q_1}\\
 &-\frac{{e_2} {q_1} \left({q_1}^{2}+{q_2}^{2}+1\right) \left(2 {a_2} \,{q_2}^{2}
 a_1(q_1^2 - q_2^2 - 1)
 \right)}{4 {e_1} \left({q_1}^{2}+{q_2}^{2}\right)}
   \end{array}
  \end{equation}
 and 
 \begin{align}\label{y-s1}
 \{q_1,f_4\}_1=&\frac{\left({q_1}^{2}+{q_2}^{2}+1\right)^{2} {p_{q_1}}}{4}+\frac{{e_2} {q_2} \left(\left(q_1^2+q_2^2\right)^2-1\right)}{4{e_1} \left( {q_1}^{2}+ {q_2}^{2}\right)}\,,
 \\
 \{q_2,f_4\}_1=&\frac{\left({q_1}^{2}+{q_2}^{2}+1\right)^{2} {p_{q_2}}}{4}-\frac{{e_2} {q_1} \left(
 \left(q_1^{2}+{q_2}^{2}\right)^2-1\right)}{4{e_1} \left( {q_1}^{2}+ {q_2}^{2}\right)}\,.
 \nonumber
 \end{align}
We have to solve these rather complicated differential equations  by using an appropriate symplectic integrator. As an example we could try to modify the Newton–St\'{o}rmer–Verlet–leapfrog method to this purpose \cite{hai03}.

According to \cite{ts06} the following  Lie-Darboux transformation $\varphi_S$  define Darboux coordinates on
 the symplectic leaves of the second linear Poisson bivector $P_2$ (\ref{p2}) 
 \begin{align*}
 {M_1}=& -\frac{\sqrt{a_2a_3}}{4}\, \left(p_{q_1}^2+q_1^{2}+\frac{4w_1}{q_1^{2}}
 +p_{q_2}^{2}+q_2^{2}+\frac{4w_2}{q_2^{2}}\right)\,,\\
 {M_2}=&\phantom{-} \frac{\sqrt{-a_1a_3}}{4}\,\left(p_{q_1}^{2}-q_1^{2}+\frac{{4w_1}}{{q_1}^{2}}+p_{q_2}^{2}
 -q_2^{2} +\frac{{4w_2}}{{q_2}^{2}}\right)\,,\\
 M_3 =&\phantom{-}\frac{ \sqrt{-a_1a_2}}{2}(q_1p_1+ q_2p_2)\,,
 \end{align*}
 and
 \begin{align*}
  {p_1}= &\frac{\sqrt{-a_1}}{4}\, \left(p_{q_1}^{2}+{q_1}^{2}+\frac{{4w_1}}{{q_1}^{2}}-p_{q_2}^{2}
  -q_2^{2}-\frac{{4w_2}}{{q_2}^{2}}\right)\,,\\
 {p_2}=&\frac{\sqrt{a_2}}{4}\, \left(p_{q_1}^{2}-q_1^{2}+\frac{{4w_1}}{{q_1}^{2}}-p_{q_2}^{2}
 +q_2^{2}-\frac{{4w_2}}{{q_2}^{2}}\right)\,,\\
 {p_3}=& \frac{\sqrt{a_3}}{2} \left({q_1} p_{q_1}-{q_2} {p_{q_2}}\right)\,.
 \end{align*}
The corresponding Casimir functions are equal to
\[
 f_2=\sqrt{-a_1a_2a_3}(w_1 - w_2)\qquad\mbox{and}\qquad
 f_3=a_1a_2a_3(w_1+w_2)\,.
\]
The motions on these symplectic sheets are rather complicated differential equations, like (\ref{x-s1}) and (\ref{x-s1}).

There are no known Darboux variables on the symplectic leaves of the rational  and cubic Poisson bivectors $P_3,\ldots,P_6$.
  
In general, the Lie-Darboux transformation $ \varphi$ exists only local. Consequently, it would be necessary to define a number of transformations in overlapping domains and ensure the smoothness between them. The performance of the network-based Lie-Darboux function in such circumstances is not clear.

If an absolutely accurate Lie-Darboux transformation coupled with a symplectic integrator is used to transform the coordinates to canonical form, it will of course preserve all the Casimir functions. However, any numerical errors in determining that transformation will result in corresponding errors in the Casimir function evolution.

As the errors in Casimir evolution are determined by the errors in the Lie-Darboux mapping, it is also not clear how these errors will accumulate over the long-term evolution of the system, see discussion in \cite{cos23, cos23a, put24, poi22}. 

To avoid these problems, it is necessary to stop trying to reduce the problem of integrating  equations of motion on Poisson manifolds to the problem of integrating  equations of motion on symplectic manifolds. In the next section, we will talk about how to do this.

\section{Kahan discretization}
 \setcounter{equation}{0}
 A family of unconventional integrators for ordinary differential equations with polynomial vector fields was recently proposed, based on the polarization of vector fields \cite{sur24}. The simplest instance is the now-famous Kahan discretization for quadratic vector fields. It has been demonstrated that all these integrators possess remarkable conservation properties. In particular, it has been proved that, when the underlying ordinary differential equations (\ref{m-eq}) is Hamiltonian, its polarization discretization possesses an integral of motion and an invariant volume form \cite{kah19,kah22,sur09,sur18}.

In his unpublished lectures \cite{kah93} Kahan introduced a numerical method designed for quadratic systems of ordinary differential equations
\[
\frac{d x_i}{d t} = \sum_{j,k=1}^n a_{ijk} x_j x_k + \sum_{j=1}^n b_{ij} x_j + c_i,\quad i=1,\ldots,n,
\]
which are replaced by the discrete equations   
\begin{equation}\label{k-dis}
   \frac{\widetilde{x}_i-x_i}{h} = \sum_{j,k} a_{ijk} \frac{\widetilde{x}_jx_k+x_j\widetilde{x}_k}{2} + \sum_j  b_{ij}\frac{x_j+\widetilde{x}_j}{2} + c_i,\quad i=1,\ldots,n.
\end{equation}
Here $a_{ijk}, b_{ij}, c_i$ are arbitrary constants, $h$ denotes the discrete time step and $\widetilde{x}=x(t+h)$. 
The method of Kahan is the restriction of a Runge-Kutta method to
quadratic vector fields which is otherwise referred to as the Hirota-Kimura discretisation \cite{hk00}. 

This discretization scheme is linearly implicit; consequently, its inverse is also implicit. Thus, it defines a birational map 
\[ \widetilde{x}=\phi(x)\,.\] 
The Kahan discretization is known to inherit first integrals and measures of the underlying quadratic differential
equation more frequently than could be anticipated, see \cite{kah19,kah22,sur09,sur18}   and a more recent literature.

It would be important to understand whether an analogue of the invariance equation for the continuous flows
\[
\mathcal L_X\,T=0
\]
in the discrete situation. The map $\phi$ has a  scalar invariant if there exists a smooth function  $f(x)$ so that
\[
f(\widetilde{x})=f(x)\,.
\]
The map $\phi$ is said to be measure-preserving if there exists a smooth function $\mu$ such that 
\[\mu(\widetilde{x})J(x) = \mu(x),\] where $J(x)$ denotes the Jacobian of $\phi$ 
 \begin{equation}
 \label{jac}
 J = \left| \frac{\partial \phi_i}{\partial x_j} \right|\,.
 \end{equation} 
In this case map $\phi$ has an invariant volume form $\Omega=\mu(x)\, dx_1 \wedge \dots \wedge dx_n$.

A polynomial $D(x)$ is  a Darboux polynomial of the map $\phi$ if there exists function $c(x)$ such that  
\begin{equation}\label{cofeqn}
D(\widetilde{x}) = c(x) D(x)\,. 
\end{equation} 
Suppose that we know few solutions of the various cofactor equations 
\[D_i(\widetilde{x}) = c_i(x) D_i(x), \qquad i=1,\dots,k,\] then we have
\[
\left( \prod_{i=1}^{k} D_i^{\alpha_i}(\widetilde{x}) \right) = \left( \prod_{i=1}^{k} c_i^{\alpha_i}(x) \right) \left( \prod_{i=1}^{k} D_i^{\alpha_i}(x)  \right). 
\]
If 
\begin{equation}\label{j-fact}
	 \prod_{i=1}^{k} c_i^{\alpha_i}(x) = J(x),
\end{equation} 
then rational function
\begin{equation}\label{d-mu}
	 \mu^{-1}(x)=\prod_{i=1}^{k} D_i^{\alpha_i}(x),
\end{equation}
defines the invariant volume form $\Omega$. As a result, factorization of the Jacobian $J$ (\ref{j-fact}) is a first step to computing cofactors $c_i(x)$  in the discrete case \cite{kah19,kah22} and in the continuous case \cite{ts25}.

If we know cofactors $c_i(x)$ in (\ref{j-fact}), we can substitute the polynomial $D_i(x) $ with undetermined coefficients in the corresponding cofactor equation (\ref{cofeqn}), solve the resulting system of linear algebraic equations and compute this Darboux polynomial $D_i(x)$. After that we can construct rational function
\begin{equation}\label{f-int}
	 f(x)=\prod_{i=1}^{k} D_i^{\beta_i}(x),
\end{equation} 
with exponents $\beta_i$ such that 
\[
	 \prod_{i=1}^{k} c_i^{\beta_i}(x) = 1.
\]
This function $f(x)$ is a first integral of the map $\phi$.

\subsection{Euler top} 
Let us consider the Euler equations describing the motion of a rigid body
\begin{equation}\label{c-eul}
\dot{x}_1=(a_2 - a_3)x_2x_3\,,\qquad \dot{x}_2=(a_3 - a_1)x_3x_1\,,\qquad \dot{x}_3=(a_1 - a_2)x_1x_2\,,
\end{equation}
where $x=(x_1,x_2,x_3)$ are the coordinates on the state space of this autonomous system of differential equations and $a_1,a_2$ and $a_3$ are arbitrary real parameters \cite{bm}. Cyclic permutation of the indexes 
\begin{equation}\label{point-sym}
(1,2,3) \to (2,3,1)\to (3,1,2)
\end{equation} is a point symmetry of the Euler equations. 

Solving the invariance equation (\ref{g-inv}) in the space of polynomials with undetermined coefficients  we obtain  two base 
scalar solutions
\[f_1=\frac12\left(a_1x_1^2+a_2x_2^2+a_3x_3^2\right)\qquad\mbox{and}\qquad f_2=\frac12\left(x_1^2+x_2^2+x_3^2\right)\,.\]
The divergence of the vector field $X$ defined by Euler equations  (\ref{c-eul}) is zero and, therefore, in the space of differential forms of type $(0,3)$ the basic solution of the invariance equation (\ref{g-inv}) is a volume form
\[
\Omega=dx_1\wedge dx_2\wedge dx_3\,,
\]
 and the basic solution in the space of multivector fields  of type (3,0) has the form
\begin{equation}\label{lcv-eul}
\mathcal E =\partial_1\wedge \partial_2 \wedge \partial_3\,.
\end{equation}
The products of the invariant  trivector $\mathcal E$ and the invariant 1-forms $df_1$ and $df_2$ are compatible Poisson bivectors on the Lie-Poisson algebra $so^*(3)$ 
\[
P_1=\mathcal E df_1=\left(
                      \begin{array}{ccc}
                        0 & a_3x_3 & -a_2x_2 \\
                        -a_3x_3 & 0 & a_1x_1 \\
                        a_2x_2 & -a_1x_1 & 0 \\
                      \end{array}
                    \right)\qquad\mbox{and}\qquad
 P_2=\mathcal E df_2=\left(
                      \begin{array}{ccc}
                        0 & x_3 & -x_2 \\
                        -x_3 & 0 & x_1 \\
                        x_2 & -x_1 & 0 \\
                      \end{array}
                    \right) \,.                  
\]
Solving equations $f_1=e_1/2$ and $f_2=e_2/2$ for $x_{1,2}$ we obtain
\[
x_1 = \sqrt{\frac{(a_3-a_2)x_3^2+ a_2e_2 - e_1}{a_2 - a_1}},\qquad 
x_2 = \sqrt{\frac{(a_1- a_3)x_3^2 - a_1e_2 + e_1}{a_2 - a_1}}\,,\qquad e_{1,2}\in\mathbb R\,.
\]
Substituting these expressions in the third Euler equation one gets equation
\begin{equation}\label{g-ell}
\Gamma:\qquad \left(\frac{dx_3}{dt}\right)^2 = \left(\left(a_3-a_2\right)x_3^2 + a_2e_2 - e_1\right)\left(\left(a_1- a_3\right)x_3^2 - a_1e_2 + e_1\right)
\end{equation}
which defined an elliptic curve $\Gamma$ with the well-known group structure. Using addition and multiplication by integer of the points on this curve, we can construct discretizations which preserve all base invariants $f_{1,2}$, $\Omega$ and $\mathcal E$ together with the invariant Poisson bivectors $P_{1,2}=\mathcal E df_{1,2}$ \cite{fed05,ts18,ts18a}.

Multiplication by integer is inseparable  isogeny of the elliptic curve $\Gamma$. Other isogenies also give rise to various  discretizations. As an example, the Kahan discretization \cite{hk00}  of the Euler equations (\ref{c-eul}) looks like
\begin{equation}\label{d-eul}
\begin{array}{l}
\widetilde{x}_1-x_1=\epsilon b_1(\widetilde{x}_2x_3+x_2\widetilde{x}_3),\\
\widetilde{x}_2-x_2=\epsilon b_2( \widetilde{x}_3x_1+x_3\widetilde{x}_1),\\
\widetilde{x}_3-x_3=\epsilon b_3 (\widetilde{x}_1x_2+x_1\widetilde{x}_2),
\end{array}
\end{equation}
 where we put $\epsilon=h/2$ and 
\[b_1= (a_2-a_3)\,,\qquad b_2=(a_3 - a_1)\,,\qquad b_3=(a_1 - a_2)\,.\]
Solving (\ref{d-eul}) for $\widetilde{x}$ we determine the map 
$\phi:\, x\mapsto{\widetilde{x}}$   which is given by
\[
\widetilde{x} =\phi(x)=A^{-1}(x)x, \qquad
A(x)=
\begin{pmatrix}
1 & -\epsilon b_1 x_3 & -\epsilon b_1 x_2 \\
-\epsilon b_2 x_3 & 1 & -\epsilon b_2 x_1 \\
-\epsilon b_3 x_2 & -\epsilon b_3 x_1 & 1
\end{pmatrix} .
\]
The Jacobian (\ref{jac}) of this map is equal to 
\[
J(x)=\frac{n^2(x)}{d^4(x)}\,,
\]
where
\begin{align*}
n(x)=1 - 2&(b_1b_2x_3^2 +b_1b_3x_2^2 +b_2b_3x_1^2) \epsilon^2 - 8b_1b_2b_3x_1x_2x_3\epsilon^3\\
   + &(b_1^2 b_2^2 x_3^4 - 2 b_1^2 b_2 b_3 x_2^2 x_3^2 + b_1^2 b_3^2 x_2^4 - 2 b_1 b_2^2 b_3 x_1^2 x_3^2 - 2 b_1 b_2 b_3^2 x_1^2 x_2^2 + b_2^2 b_3^2 x_1^4) \epsilon^4\,,\\ \\
d(x)=1 -\phantom{2} &(b_1b_2x_3^2 + b_1b_3x_2^2 + b_2b_3x_1^2)\epsilon^2- 2b_1b_2b_3x_1x_2x_3\epsilon^3   \,. 
\end{align*}
Substituting cofactor
\[
c(x)=\frac{n(x)}{d^2(x)}\,,\qquad  c^2(x)=J(x)\,,
\]
and polynomial of second order 
\[
D(x)=\sum_{i>j}^3 u_{ij}x_ix_j+\sum u_jx_i+u_0, \qquad u_{ij},u_j,u_0\in\mathbb R
\]
in the cofactor equation (\ref{cofeqn}) we obtain a system of linear algebraic equations for undetermined coefficients 
 $u_{ij},u_j$ and $u_0$. The generic solution of this system is a linear combination of the six  Darboux polynomials
 \[
 D_1(x)=b_1(1-\epsilon^2b_2b_3x_1^2)\,,\quad D_2(x)=b_2(1-\epsilon^2b_3b_1x_2^2)\,,\quad
 D_3(x)=b_3(1-\epsilon^2b_1b_2x_3^2)\,,
 \]
 and
 \[
 D_4=(b_1x_2^2 - b_2x_1^2)\,,\quad D_5=(b_2x_3^2 - b_3x_2^2)\,,\quad D_6= (b_3x_1^2 - b_1x_2^2)
 \]
 related by the point symmetry (\ref{point-sym}). It allows us to construct first integrals (\ref{f-int})
 \[
 f_{ij}=D_iD^{-1}_j
 \]
invariant volume forms (\ref{d-mu})
\[
\Omega_{ij}=(D_iD_j)^{-1} \,dx_1\wedge dx_2\wedge dx_3\]
 and invariant trivectors
 \[ \mathcal E_{ij}=D_iD_j\,\partial_1\wedge\partial_1\wedge\partial_3\,,
\] 
The tensor products of invariant trivectors $\mathcal E_{ij}$ with invariant 1-forms $df_{km}$ are invariant bivectors
\[
P_{ijkm}=\mathcal E_{ij}df_{km}\,.
\]
These can be combined to invariant Poisson bivectors that satisfy the Poisson condition \cite{sur16}.    

Solve two independent first integrals for $x_{1,2}$ and substitute these expressions into
\[\tilde{\Gamma}:\quad \widetilde{x}_3=\phi_3(x_1,x_2,x_3)\]
we obtain an equation defining elliptic curve $\tilde{\Gamma}$ \cite{sur16}, which is related by isogeny to the elliptic curve $\Gamma$ (\ref{g-ell}) in the continuous case.

\subsection{Clebsch system}
For the Clebsch system equations of motion on symplectic leaves are differ form the standard Hamiltonian system with two degrees of freedom, see (\ref{x-s1}) and (\ref{y-s1}). So, we have to modify the usual symplectic integrators and again to prove reversibility, symplecticity, volume preservation, and conservation of first integrals for these geometric integrators.

The initial equations of motion (\ref{m-eq}) defined by (\ref{x-cl}) and (\ref{y-cl}) are simpler than equations on symplectic leaves. Applying the Kahan scheme (\ref{k-dis}) to the vector field $X$ (\ref{x-cl}), we directly arrive at a birational map $\phi_X:\,(M,p)\mapsto (\widetilde{M},\widetilde{p})$, defined by the following linear system
\begin{equation}\label{x-dcl}
 \left\{ \begin{array}{l}
\widetilde{M}_1-M_1 =
\epsilon(a_3-a_2)(\widetilde{M}_2M_3+M_2\widetilde{M}_3)+
\epsilon a_1(a_3-a_2)(\widetilde{p}_2p_3+p_2\widetilde{p}_3),
\vspace{.1truecm} \\
\widetilde{M}_2-M_2 =
\epsilon(a_1-a_3)(\widetilde{M}_3M_1+M_3\widetilde{M}_1)+
\epsilon a_2(a_1-a_3)(\widetilde{p}_3p_1+p_3\widetilde{p}_1),
\vspace{.1truecm}\\
\widetilde{M}_3-M_3 =
\epsilon(a_2-a_1)(\widetilde{M}_1M_2+M_1\widetilde{M}_2)+
\epsilon a_3(a_2-a_1)(\widetilde{p}_1p_2+p_1\widetilde{p}_2),
\vspace{.1truecm} \\
\widetilde{p}_1-p_1 = \epsilon
a_3(\widetilde{M}_3p_2+M_3\widetilde{p}_2)-\epsilon
a_2(\widetilde{M}_2 p_3+M_2\widetilde{p}_3),
\vspace{.1truecm}\\
\widetilde{p}_2-p_2= \epsilon
a_1(\widetilde{M}_1p_3+M_1\widetilde{p}_3)-\epsilon
a_3(\widetilde{M}_3p_1+M_3\widetilde{p}_1),
\vspace{.1truecm} \\
\widetilde{p}_3-p_3 = \epsilon
a_2(\widetilde{M}_2p_1+M_2\widetilde{p}_1)-\epsilon
a_1(\widetilde{M}_1 p_2+M_1\widetilde{p}_2).
\end{array}\right.
\end{equation}
 where we put $\epsilon=h/2$ following to \cite{sur09,sur18}. 

The Kahan discretization $\phi_Y:\,(M,p)\mapsto (\widetilde{M},\widetilde{p})$ of the invariant vector field $Y$ (\ref{y-cl}) looks like
\begin{equation} \label{y-dcl}
\left\{\begin{array}{l}
\widetilde{M}_1-M_1  =  \epsilon(a_3-a_2)
(\widetilde{p}_2p_3+p_2\widetilde{p}_3),         \vspace{.1truecm} \\
\widetilde{M}_2-M_2  =  \epsilon(a_1-a_3)
(\widetilde{p}_3p_1+p_3\widetilde{p}_1),         \vspace{.1truecm} \\
\widetilde{M}_3-M_3  =  \epsilon(a_2-a_1)
(\widetilde{p}_1p_2+p_1\widetilde{p}_2),         \vspace{.1truecm} \\
\widetilde{p}_1-p_1  = 
\epsilon(\widetilde{M}_3p_2+M_3\widetilde{p}_2)-
\epsilon(\widetilde{M}_2p_3+M_2\widetilde{p}_3), \vspace{.1truecm} \\
\widetilde{p}_2-p_2  = 
\epsilon(\widetilde{M}_1p_3+M_1\widetilde{p}_3)-
\epsilon(\widetilde{M}_3p_1+M_3\widetilde{p}_1),  \vspace{.1truecm} \\
\widetilde{p}_3-p_3  = 
\epsilon(\widetilde{M}_2p_1+M_2\widetilde{p}_1)-
\epsilon(\widetilde{M}_1p_2+M_1\widetilde{p}_2). 
\end{array}\right.
\end{equation}
Associated with the commuting vector fields $X$ (\ref{x-cl}) and $Y$ (\ref{y-cl}) maps $\phi_X$ (\ref{x-dcl}) and $\phi_Y$ (\ref{y-dcl}) do not commute with each other \cite{sur09}. As a result, we have two different integrable maps with different first integrals  and invariant volume forms, which can be found in \cite{sur18}.  Nothing is known about the existence of invariant Poisson structures for these maps.

First integrals and invariant volume forms for discrete  maps $\phi_X$ (\ref{x-dcl}) and $\phi_Y$ (\ref{y-dcl}) were obtained in \cite{sur09,sur18} using  computer algebra systems. These results can be recover in the framework of the brute-force Darboux  method which consists  of the following main steps
\begin{enumerate}
\item Compute the Jacobian determinant of the map that is analysed;
\item Factor the Jacobian determinant;
\item Calculate the corresponding Darboux polynomials.
\end{enumerate}
 According to \cite{kah19,kah22}, the bottleneck for such computations often seems to be memory consumption rather than computational complexity. For the Clebsch system, the main problem is a factorisation of the numerator of the Jacobian, which requires the use of modern methods of neural networks. A discussion of these computations is beyond the scope of this paper, as is Lagutinsky's approach to finding Darboux polynomials \cite{lag1,lag2} in the discrete case. 
 
The second problem is the presentation of the results of computer analysis, which are very bulky. In \cite{sur09,sur18}, the authors solved this problem by using functions $f(x,\widetilde{x})$ instead of $f(x)$. The statement that function $f(x,\widetilde{x})$ is a conserved quantity of $\phi$  means
that
\[f(x, \phi(x)) = f(\phi(x), \phi^2(x))\,. \]
We will not give here the results of the works \cite{sur09,sur18} which were obtained by the introduction of special bases in the polynomial ring.

 \section{Conclusion}
The subject of geometric numerical integration deals with numerical integrators
that preserve geometric properties of the flow of a differential equation,
and it explains how structure preservation leads to the excellent
long-time behaviour of these integrators: long-time energy conservation, linear
error growth and preservation of invariant tori in near-integrable systems,
a discrete virial theorem, and preservation of adiabatic invariants, etc.

Tensor invariants continuous differential equations satisfy to the invariance equation (\ref{g-inv})
\[\mathcal L_X T=0\,.\]
In this note, we solved this equation for the Clebsch system using simple polynomial anzats and obtain linear, cubic and rational invariant Poisson bivectors. These geometric structures are preserved by symplectic integrators on the corresponding symplectic leaves a'priory. The main problem here is construction of the suitable Darboux coordinates on the symplectic leaves.

Using Lax matrices for the Clebsch system and Moser-Veselov refactorisation method we could obtain an explicit, second-order, integrable approximation of the continuous solutions. It will be interesting to study invariant Poisson structures for this discretization and their relation with classical $r$-matrices.

Applying the Kahan scheme  to the Clebsch equations of motion  we  obtain  well-studied birational map which
approximates the flow of the governing continuous differential equations. Nothing is known about the existence of invariant Poisson structures for this integrable map. Furthermore, an analog of the invariance equation for discrete systems has not been identified in the extant literature even for the Kahan map.

According to \cite{bob86,ts06} there is an isomorphism between the Clebsch system and the Frahm-Schottky system. This allows to directly reproduce all presented results for this system as well. Similar tensor invariants also exist for the Steklov-Lyapunov system, for the integrable and nonintegrable nonholonomic Suslov problem and some other classical systems integrable by the Euler-Jacobi theorem \cite{ts25}.

This work of  A.V. Tsiganov was supported by the Russian Science Foundation under grant no. 19-71-30012, https://www.rscf.ru/project/23-71-33002/ and performed at the Steklov Mathematical Institute of the Russian Academy of Sciences.

\end{document}